\documentclass[preprint,12pt]{elsarticle}
\journal{Journal of Subatomic Particles and Cosmology}

 \biboptions{comma,sort&compress}

\usepackage[table]{xcolor}

\usepackage{graphicx}
\usepackage{amsmath}
\usepackage{here}
\usepackage{cuted}

\usepackage{here}
\usepackage[dvips]{epsfig}
\definecolor{bluette}{rgb}{.2,.4,0}
\definecolor{salmon}{rgb}{.9,0.68,0.5}
\definecolor{motive}{rgb}{0.2,1,.5}
\definecolor{list}{rgb}{0.3,.8,.1}
\definecolor{moe}{rgb}{1,.7,.5}
\definecolor{mote}{rgb}{.7,.5,.6}
\definecolor{pisello}{rgb}{.1,1,0}
\definecolor{orange}{rgb}{1,.7,0}
\definecolor{oliva}{rgb}{.1,.5,0.3}
\definecolor{greenda}{rgb}{0,.3,.2}
\definecolor{greenli}{rgb}{0.5,.8,.0}
\definecolor{blueda}{rgb}{0,.1,.6}
\definecolor{purple}{rgb}{.7,.1,.2}
\definecolor{marrone}{rgb}{1,0.7,0}
\definecolor{pinky}{rgb}{1,0.8,0.8}
\definecolor{rose}{rgb}{1,0.4,0}

\def\oliva{\color{oliva}}

\def\beq{\begin{equation}}
\def\eeq{\end{equation}}
\def\bea{\begin{eqnarray}}
\def\eea{\end{eqnarray}}
\def\bq{\begin{quote}}
\def\eq{\end{quote}}

\def\nnb{\nonumber}
\def\ga{\left(}
\def\dr{\right)}

\def\lrar{\Longrightarrow}
\def\lrar2{\longrightarrow}
																
\def\nnb{\nonumber}

\def\la{\langle}
\def\ra{\rangle}

\def\ba{\vspace*{-0.2cm}\begin{array}}
\def\ea{\end{array}\vspace*{-0.2cm}}

\def\b{$\bullet~$}
\def\d{$\diamond~$}

\def\als{\alpha_s}

\def\gg2{\la\alpha_s G^2 \ra}
\def\gg3{g^3f_{abc}\la G^aG^bG^c \ra}
\def\ggg4{\la\als^2G^4\ra}

\def\gg{\lag g^{2}_{s} G^2 \rag}
\def\ggg{\lag g^{3}_{s}G^3\rag}

\usepackage{amsmath}
\usepackage{slashed}
\usepackage{color}

\begin{document}
\begin{frontmatter}

\title{QCD condensates and $\alpha_s$  from $\tau$-decay: Summary\,\tnoteref{invit}}
\tnotetext[invit]{Talk given at the 14th International Conference in High-Energy Physics - HEPMAD24, 21-26th october 2024, Antananarivo-MG.}
\author{Stephan Narison
}
\address{Laboratoire
Univers et Particules de Montpellier (LUPM), CNRS-IN2P3, \\
Case 070, Place Eug\`ene
Bataillon, 34095 - Montpellier, France\\
and\\
Institute of High-Energy Physics of Madagascar (iHEPMAD)\\
University of Ankatso, Antananarivo 101, Madagascar}
\ead{snarison@yahoo.fr}


\date{\today}
\begin{abstract}
In this talk, I summarize the recent determinations in Ref.\,\cite{SNtau24} of the 
 QCD condensates and $\alpha_s$ within the SVZ expansion  using the ratio of Laplace sum rule (LSR) ${\cal R}_{10}^A(\tau)$  and $\tau$-like moments ${\cal R}_{n,A},\, {\cal R}_{n,V-A}$within stability criteria in the axial-vector (A) and V-A channels of the $\tau$- semi-hadronic decays.   The factorization of the four-quark condensate is violated by a factor 6 like in the case of the $e^+e^-\to$ Hadrons data. There is no exponential growth of the size of higher dimension condensates which does not favour (by duality) a significant effect of the so-called Duality Violation (DV).  The extracted value of $\alpha_s$ is in good agreement with the one from  $e^+e^-$ and indicates  that the determination  at the observed $\tau$-mass is not the optimal one and leads to an overestimate of its actual value. 
 \end{abstract}

\end{frontmatter}

\newpage
\section{Introduction}
\vspace*{-0.2cm}

A precise determination of the QCD  condensates is an important step for understanding the theory of QCD and its vacuum structure while the one of $\alpha_s$ is useful for PT QCD and as input in the precision test of the Standard Model (SM). 
In this paper,  I summarize the analysis and  the results in Ref.\,\cite{SNtau24} for the determinations of the QCD condensates and $\alpha_s$ from the Axial-Vector and V-A channels of the semi-hadronic $\tau$-decays which we shall compare with the ones from $e^+e^-\to$ I=1 Hadrons data and from some other determinations. 

The approach is similar to the one in \,\cite{SNe,SNe2} from $e^+e^-\to$ I=1 Hadrons and Vector component of $\tau$-decays. A  comparison with the corresponding results is done.

\section{The axial-vector (A)  and vector(V)--A two-point functions}
We shall be concerned with the two-point correlator :
\bea
\hspace*{-0.6cm} 
\Pi^{\mu\nu}_{V(A)}(q^2)&=&i\hspace*{-0.1cm}\int \hspace*{-0.15cm}d^4x ~e^{-iqx}\la 0\vert {\cal T} {J^\mu_{V(A)}}(x)\ga {J^\nu_{V(A)}}(0)\dr^\dagger \vert 0\ra \nnb\\
&=&-(g^{\mu\nu}q^2-q^\mu q^\nu)\Pi^{(1)}_{V(A)}(q^2)+q^\mu q^\nu \Pi^{(0)}_{V(A)}(q^2)
 \label{eq:2-point}
 \eea
built from the T-product of the bilinear axial-vector  (A) and V(vector)--A currents of $u,d$ quark fields with:
\beq
 J^\mu_{A}(x)=: \bar\psi_u\gamma^\mu\gamma_5\psi_d:,~~~~~
 J^\mu_{V-A}(x)=: \bar\psi_u\gamma^\mu(1-\gamma_5)\psi_d:,
\eeq
The upper  indices (0) and (1) correspond to the spin of the associated hadrons. The two-point function obeys the dispersion relation:
\beq
\Pi_{V(A)}(q^2)=\int_{t>}^\infty \frac{dt}{t-q^2-i\epsilon} \frac{1}{\pi}\,{\rm Im} \Pi_{V(A)}(t)+\cdots,
\eeq
where $\cdots$ are subtraction constants polynomial in $q^2$ and $t>$   is the hadronic threshold.
\subsection*{\b  The QCD two-point function within the SVZ-expansion}
According to\,\cite{SVZ}, the two-point function can be expressed in terms of the sum of higher and higher quark and gluon condensates:
\beq
4\pi^2\Pi_H(-Q^2,m_q^2,\mu)=\sum_{D=0,2,4,..}\hspace*{-0.25cm}\frac{C_{D,H}(Q^2,m_q^2,\mu)\la O_{D,H}(\mu)\ra}{(Q^2)^{D/2}}\equiv \sum_{D=0,2,4,..}\hspace*{-0.25cm}\frac {d_{D,H}} {(Q^2)^{D/2}}~, 
\label{eq:ope}
\eeq
where $H\equiv V(A)$, $\mu$ is the subtraction scale which separates the long (condensates) and short (Wilson coefficients) distance dynamics and $m_q$ is the quark mass. 

\d The non-perturbative contributions to the two-point function is given up to $d_{8,A}$ 
\,\cite{BNP,BNP2}.

\d The perturbative expression of the spectral function is known to order $\alpha_s^4$ as given explicitly in\,\cite{SNtau24,SNB2}. 
We shall use the value $\Lambda =(342\pm 8)$ MeV  for  $n_f=3$\:\, from the PDG average\,\cite{PDG}.

\subsection*{\b The A and V--A spectral functions}
\begin{figure}[H]
\begin{center}
\includegraphics[width=7.3cm]{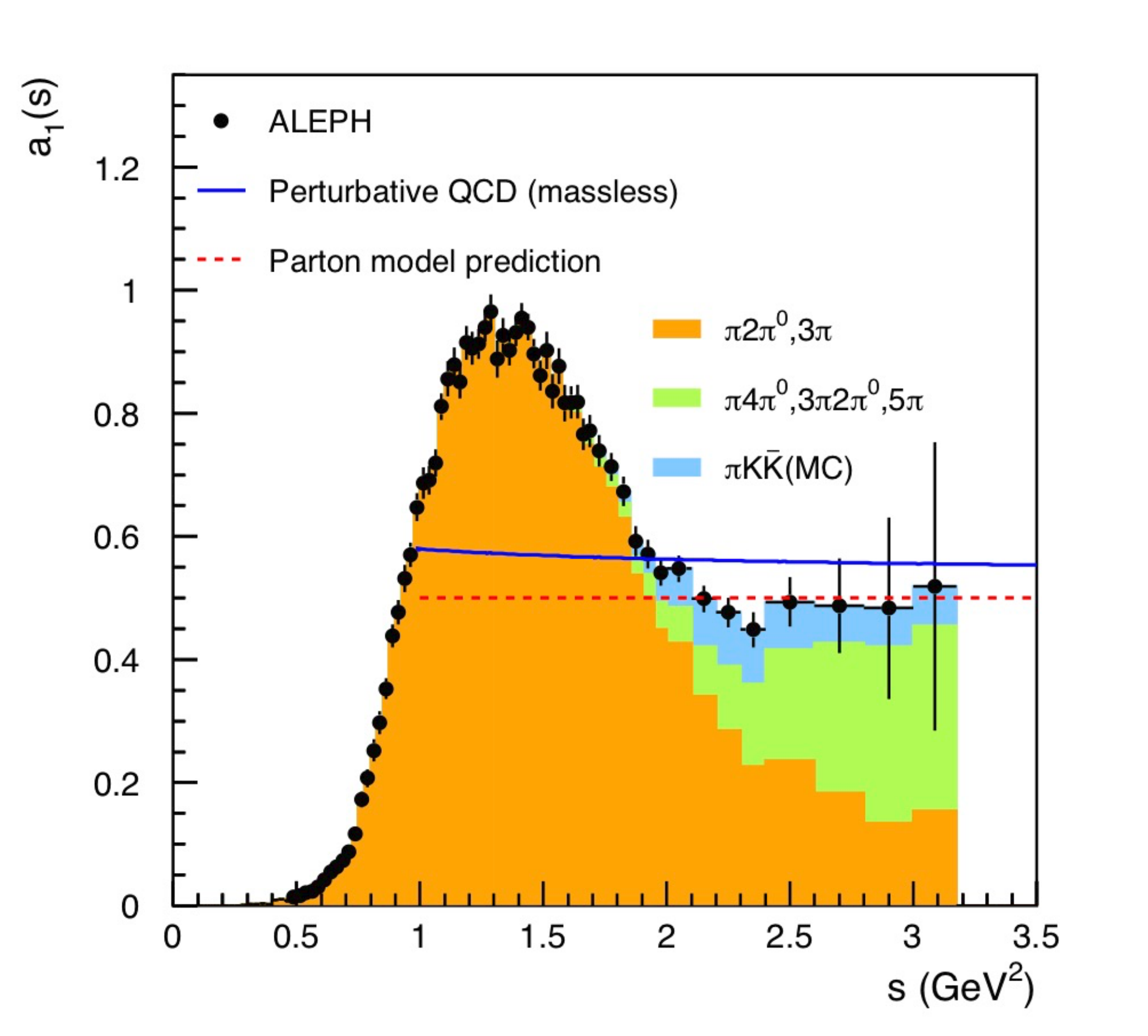}
\includegraphics[width=7cm]{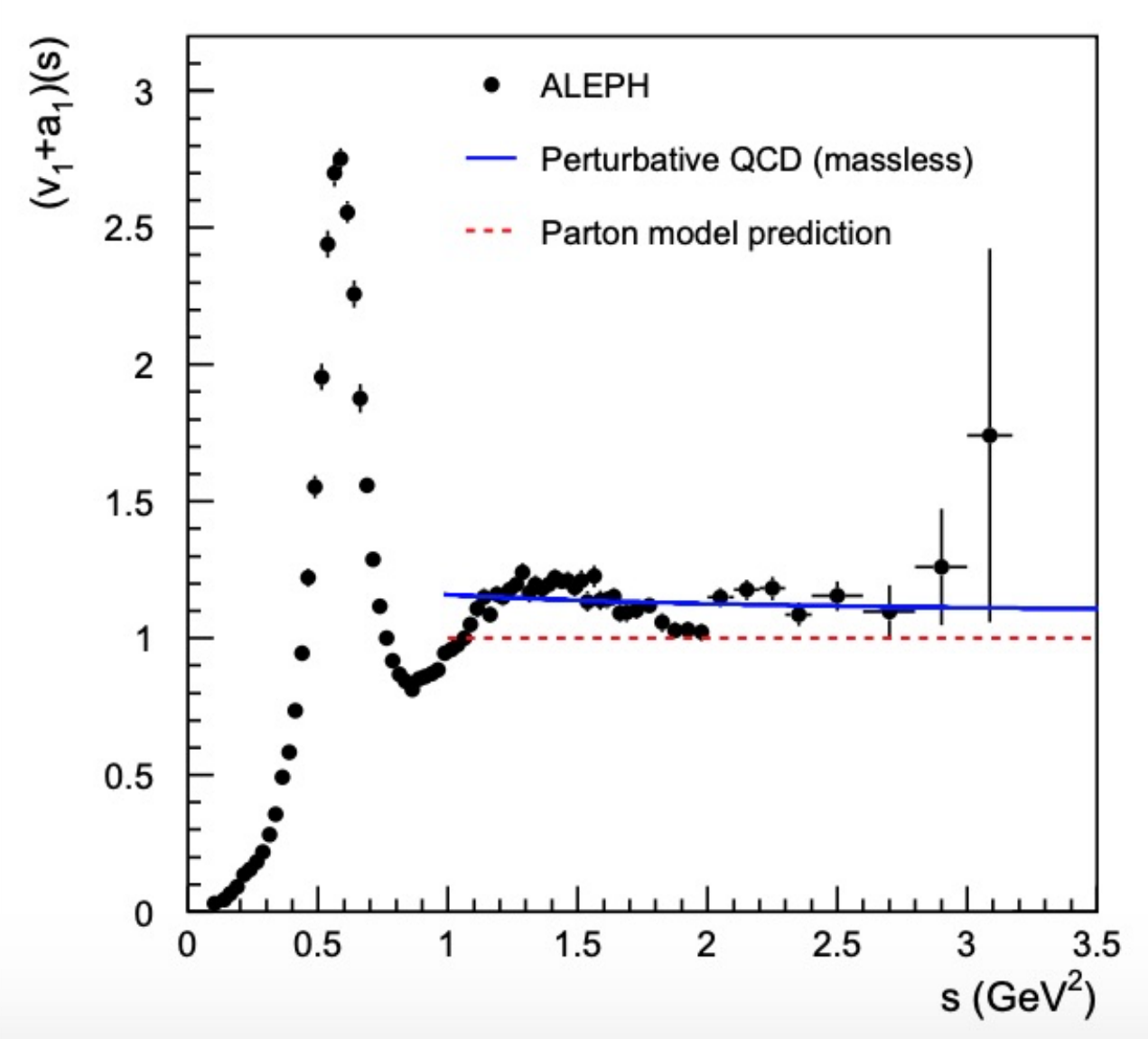}
\caption{\footnotesize  ALEPH data of the axial-vector and V--A spectral functions. } \label{fig:aleph}
\end{center}
\vspace*{-0.5cm}
\end{figure}

We shall use the recent ALEPH data\,\cite{ALEPH}  in Fig.\,\ref{fig:aleph} for the spectral function $a_1(s)$ and $v_1+a_1$ associated respectively to the A and V--A currents. 

Like in the case of the vector spectral function from $e^+e^-\to$ Hadrons\,\cite{SNe2}, we subdivide the region from threshold to the physical $\tau$ mass squared $M_\tau^2=3.16$ GeV$^2$ into different subregions 
and fit the data with 2nd and 3rd order polynomials using the optimized Mathematica program FindFit.  The details of the analysis are described in Ref.\,\cite{SNtau24}. 
\section{The ratio of Laplace sum rule (LSR) moments}
\d Like in the case of vector channel, we shall work here with the ratio of LSR moments\,\cite{SVZ,SNR,BELL}\footnote{For a recent review, see e.g.\,\cite{SNLSR}.}:
\beq
 {\cal R}^A_{10}(\tau)\equiv\frac{{\cal L}^c_{1}} {{\cal L}^c_0}= \frac{\int_{t>}^{t_c}dt~e^{-t\tau}t\, \frac{1}{\pi}\,{\rm Im}
 \Pi_H(t,\mu^2,m_q^2)}   {\int_{t>}^{t_c}dt~e^{-t\tau} \, \frac{1}{\pi}\,{\rm Im}
 \Pi_H(t,\mu^2,m_q^2)},
\label{eq:lsr}
\eeq
where

$\tau$ is the LSR variable, $t>$   is the hadronic threshold.  Here $t_c$ is  the threshold of the ``QCD continuum" which parametrizes, from the discontinuity of the Feynman diagrams, the spectral function  ${\rm Im}\,\Pi_H(t,m_q^2,\mu^2)$.  $m_q$ is the quark mass and $\mu$ is an arbitrary subtraction point.

\d The PT expression of the two-point function is known to order $\alpha_s^4$  as explicitly given in Ref.\,\cite{SNtau24}.  Within the SVZ-expansion, the non-perturbative 
contributions to the lowest LSR moment in terms of the $d_D$ condensates is\,:
\beq
{\cal L}^{NPT}_0(\tau) = \frac{3}{2}\tau^{-1}\sum_{D\geq 2}\frac{d_D}{(D/2-1)!} \tau^{D/2} ~,
\label{eq:svz}
\eeq
from which one can deduce ${\cal L}^{NPT}_1$ and ${\cal R}^A_{10}$. 
\section{Determination of the QCD condensates from $R^A_{10}$}
In so doing, we shall truncate the OPE at $d_8$
\subsection*{\b The dimension-six and -eight condensates }
\begin{figure}[hbt]
\begin{center}
\hspace*{0.cm}{\bf a)} \hspace*{6cm}{\bf b)}\\
\includegraphics[width=7.1cm]{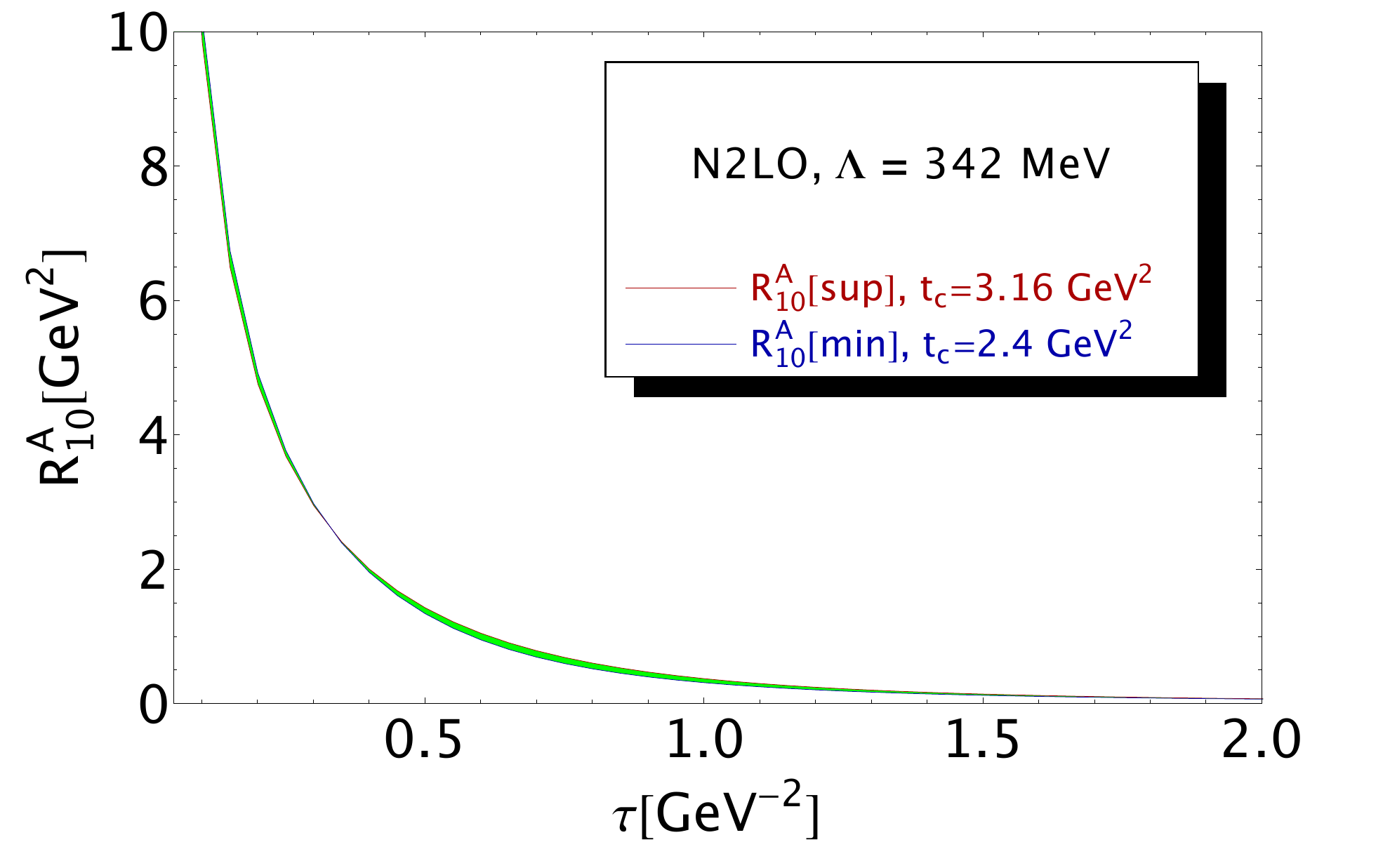}
\includegraphics[width=8.cm]{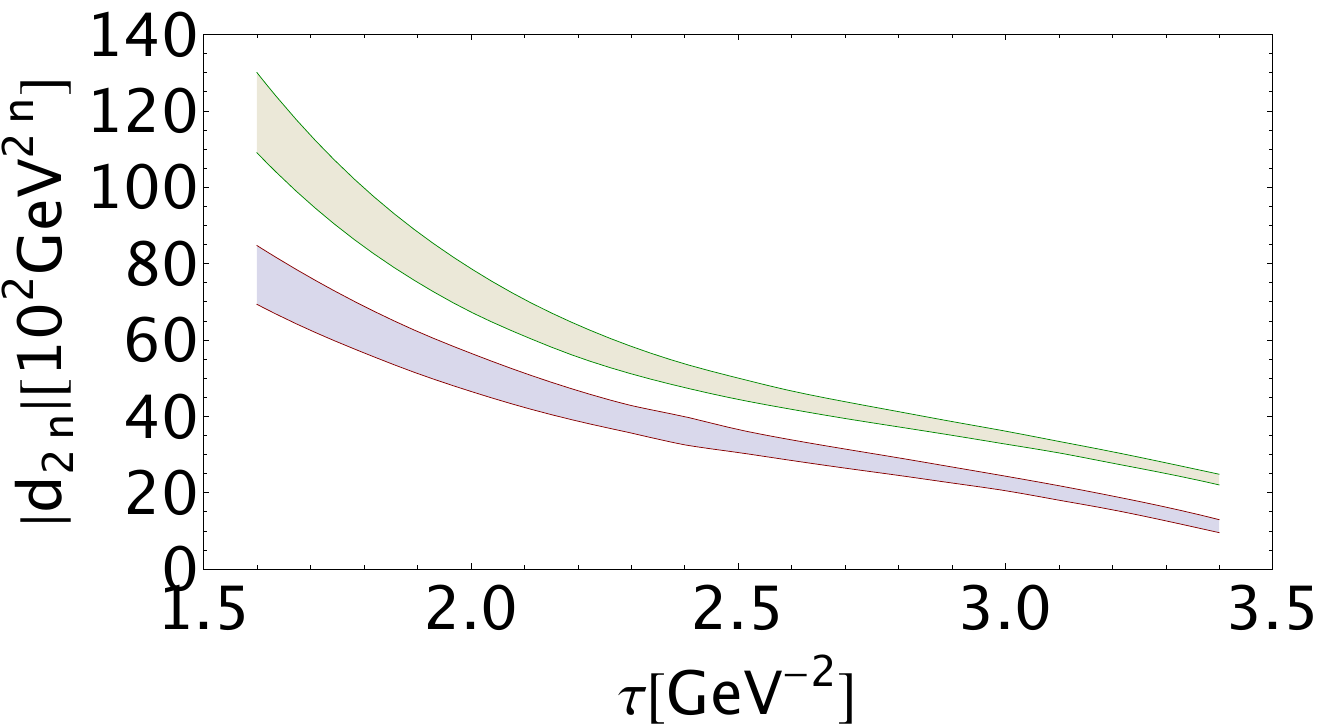}
\caption{\footnotesize {\bf a)}: $R^A_{10}$ versus the LSR variable $\tau$;  {\bf b)}:  $d_{6,A}$ and $d_{8,A}$ from the $ {\cal R}^A_{10}$} \label{fig:r10}
\end{center}
\vspace*{-0.5cm}
\end{figure}

We show in Fig.\,\ref{fig:r10}a) the $\tau$ behaviour of the phenomenological side of $ {\cal R}^A_{10}$ (experiment $\oplus$ QCD continuum beyond the threshold $t_c$) for values of $t_c$ around the physical $\tau$-lepton mass squared where the effect of $t_c$ is negligible. 
We use as input the  precise value of  $\la \alpha_s G^2\ra$  from the heavy quark mass-splittings and some other sum rules\,\cite{SNparam,SNcb1}\,\footnote{The contribution of the $\la \bar \psi\psi\ra$ condensate is multiplied by the light quark running mass $m_q$  in the OPE and can be estimated via pion PCAC as $(m_u+m_d)\la \bar\psi\psi\ra=-m_\pi^2 f_\pi^2$ where $f_\pi=93$ MeV. This contribution known to order $\alpha_s$\,\cite{BNP} is taken into account in the analysis\,\cite{SNtau24}  but is relatively negligible.}
\beq
\la\alpha_s G^2\ra =  (6.39\pm 0.35)\times 10^{-2}\,{\rm GeV^4},
\label{eq:g2}
\eeq
and perform a two-parameter fit $(d_{6,A},d_{8,A})$ by confronting the phenomenological and QCD side of $R^A_{10}$ for different values of $\tau$. 
The results of the analysis are shown in  Fig.\,\ref{fig:r10}b).  The optimal result is extracted at the inflexion point around $\tau\simeq 2.5 $ GeV$^{-2}$:
\beq
d_{6,A} =  (33.5\pm 3.0\pm 2.7)\times 10^{-2}\,{\rm GeV^6} ~ ~~~~~~~~~d_{8,A} = -(47.2\pm 2.8\pm 3.2)\times 10^{-2}\,{\rm GeV^8}
\label{eq:d68-ratio}
\eeq
where the errors come respectively from the fitting procedure and the localization of $\tau\simeq  (2.5\pm 0.1)$ GeV$^{-2}$. We expect that at this scale the OPE converges like in the case of the vector channel.
\begin{figure}[hbt]
\begin{center}
\includegraphics[width=9cm]{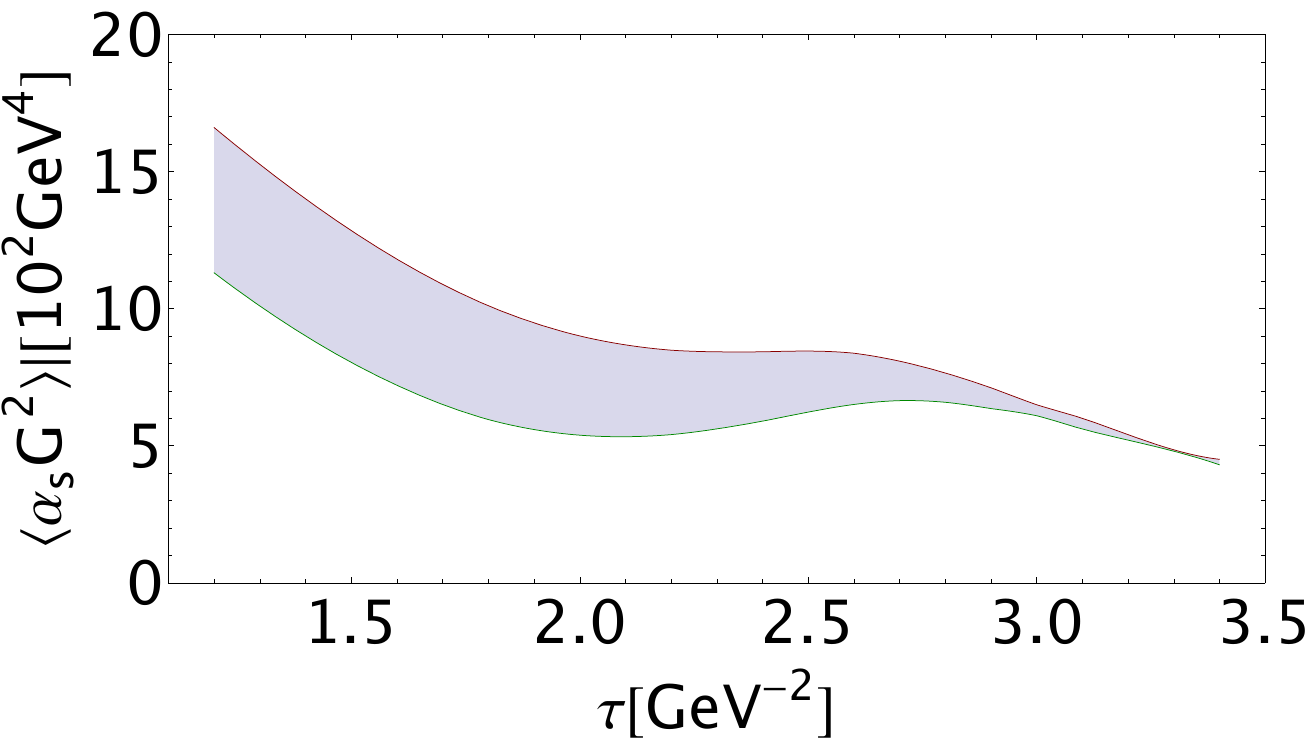}
\caption{\footnotesize  $\la \alpha_s G^2\ra$ versus the LSR variable $\tau$. } \label{fig:g2}
\end{center}
\vspace*{-0.5cm}
\end{figure} 
\subsection*{\b   The gluon condensate $\la \alpha_s G^2\ra$ from $ {\cal R}^A_{10}$}  
We use the previous values of $d_{6,A}$ and $d_{8,A}$  into $ {\cal R}^A_{10}$ and we re-extract $\la \alpha_s G^2\ra$ using a one-parameter fit.
 The analysis is shown in Fig.\,\ref{fig:g2}. We obtain:
\beq
\la \alpha_s G^2\ra =(6.9\pm 1.5)\times 10^{-2}\,{\rm GeV}^4,
\eeq
in good agreement with the one in Eq.\,\ref{eq:g2} from the heavy quark systems used previously as inputs though less accurate. This result  improves  the large range of values of the gluon condensate  extracted from $\tau$-decay moments given in the literature. It 
also shows the self-consistency of the set of condensates obtained from our analysis. 
\section{The generalized  $\tau$-decay moments}
As emphasized by BNP in Ref.\,\cite{BNP,BNP2}, it is more convenient to express the moment in terms of the combination of Spin (1+0) and Spin 0 spectral functions in order to avoid some eventual pole from $\Pi^{(0)}$ at $s=0$:
  \beq
  {\cal R}_{n,H}=6\pi\,i\,\int_{|s|=M_0^2} \hspace*{-0.5cm}dx\, (1-x)^2\,x^n\ga (1+2x) \,\Pi^{(1+0)}_H(x) -2x\Pi^{(0)}_H(x)\dr
  \eeq
  with $x\equiv s/M^2_0$, $H\equiv V,A$. $n$ indicates the degree of moment.  Some other variants of the moments have been presented in Ref.\,\cite{LEDI}. 
\section{The lowest BNP moment ${\cal R}_{0,A}$}  
It corresponds to the physical $\tau$-decay process\cite{BNP2,BNP} .

\d  Its QCD expression to order $\alpha_s^4$ and up to $d_8$ condensates
  is explicitly given in Ref.\,\cite{BNP2,BNP}.  The size of the  PT $a_s^5$ coefficient has been estimated by observing that the calculated coefficients of the  PT series grow geometrically\,\cite{SNZ}.  In the following, we consider this contribution as a source of systematic errors for the truncation of the PT series.
  
\d Using the previous fit of the ALEPH data, 
the behaviour of the lowest moment ${\cal R}_{0,A}$ versus an hypothetical $\tau$-lepton mass squared $s_0\equiv M^2_0$ is shown explicitly in Ref.\,\cite{SNtau24}. One obtains:
${\cal R}_{0,A}= 1.698(14)$ 
at the observed mass value:  $M_\tau=1.777$ GeV which is in agreement with the ALEPH value\,\cite{ALEPH} :  
${\cal R}_{0,A}\vert_{\rm Aleph}= 1.694(10)$.

\section{Beyond the SVZ-expansion\label{sec:beyond} }
 We shall not include some contributions beyond the SVZ expansion namely\,:
 
\d The $D=2$ contribution due to an eventual tachyonic gluon mass which has been introduced in Ref.\,\cite{CNZ} to parametrize phenomenologically the contribution of UV renormalon in the PT series\,\footnote{For an analysis using a large $\beta$-approximation see  e.g.\,\cite{CVETIC}.}. However, it is
expected that contribution is dual to the estimate non-calculated Higher Order terms,\cite{SNZ} which we assume to be mimiced  by our estimate of $\alpha_s^5$.
  
\d Instanton which are expected to negligible\,\cite{SNe}. 

\d Duality violation (DV)\,\cite{DV} shown in\,\cite{PICH1} to be negligible. We shall also see in the following that this contribution may be neglected. 
\section{The dimension-six and -eight condensates from ${\cal R}_{0,A}$}  

\begin{figure}[hbt]
\begin{center}
\hspace*{-4cm} {\bf a)} \hspace*{8cm}{\bf b)}\\
\includegraphics[width=7cm]{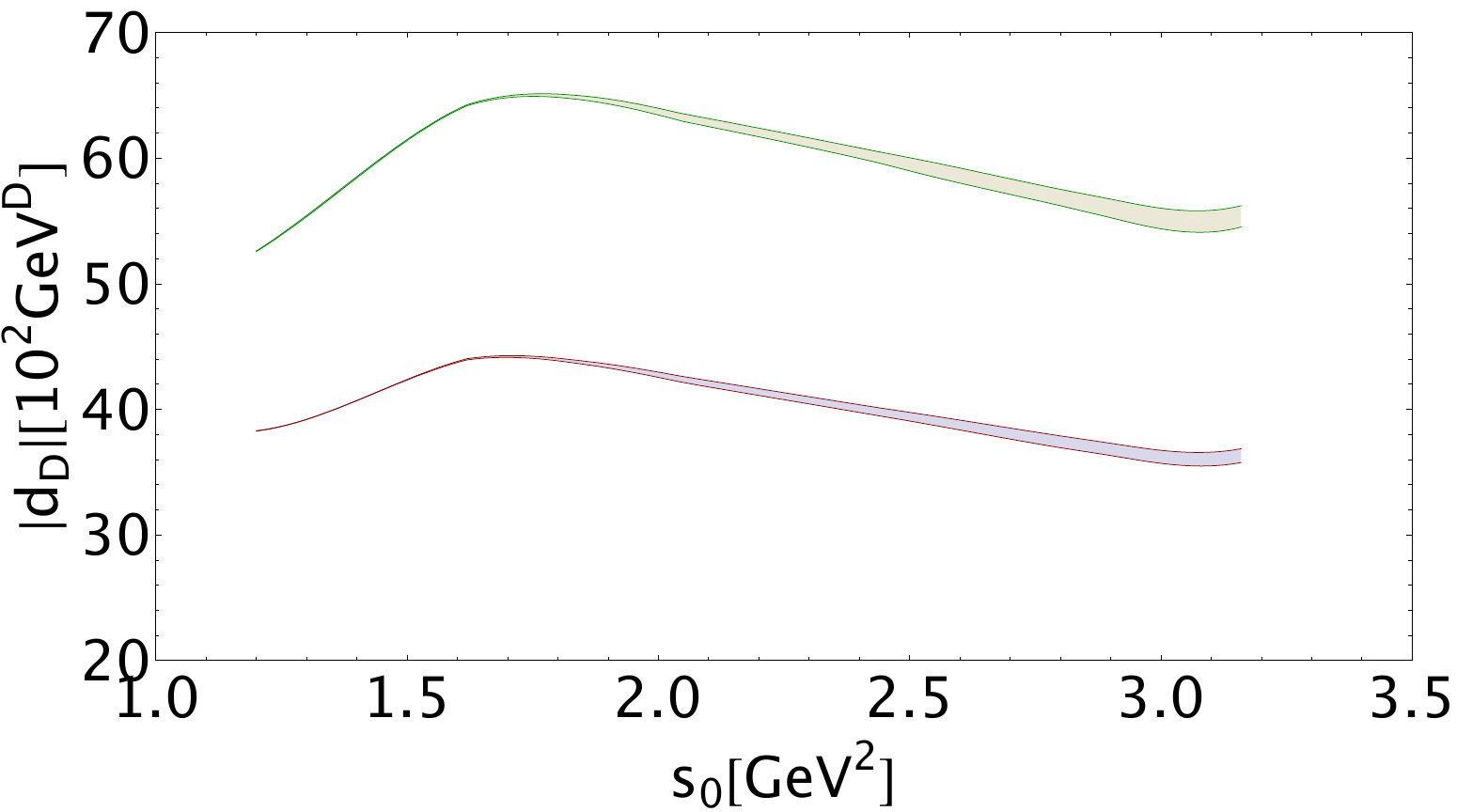}
\includegraphics[width=7cm]{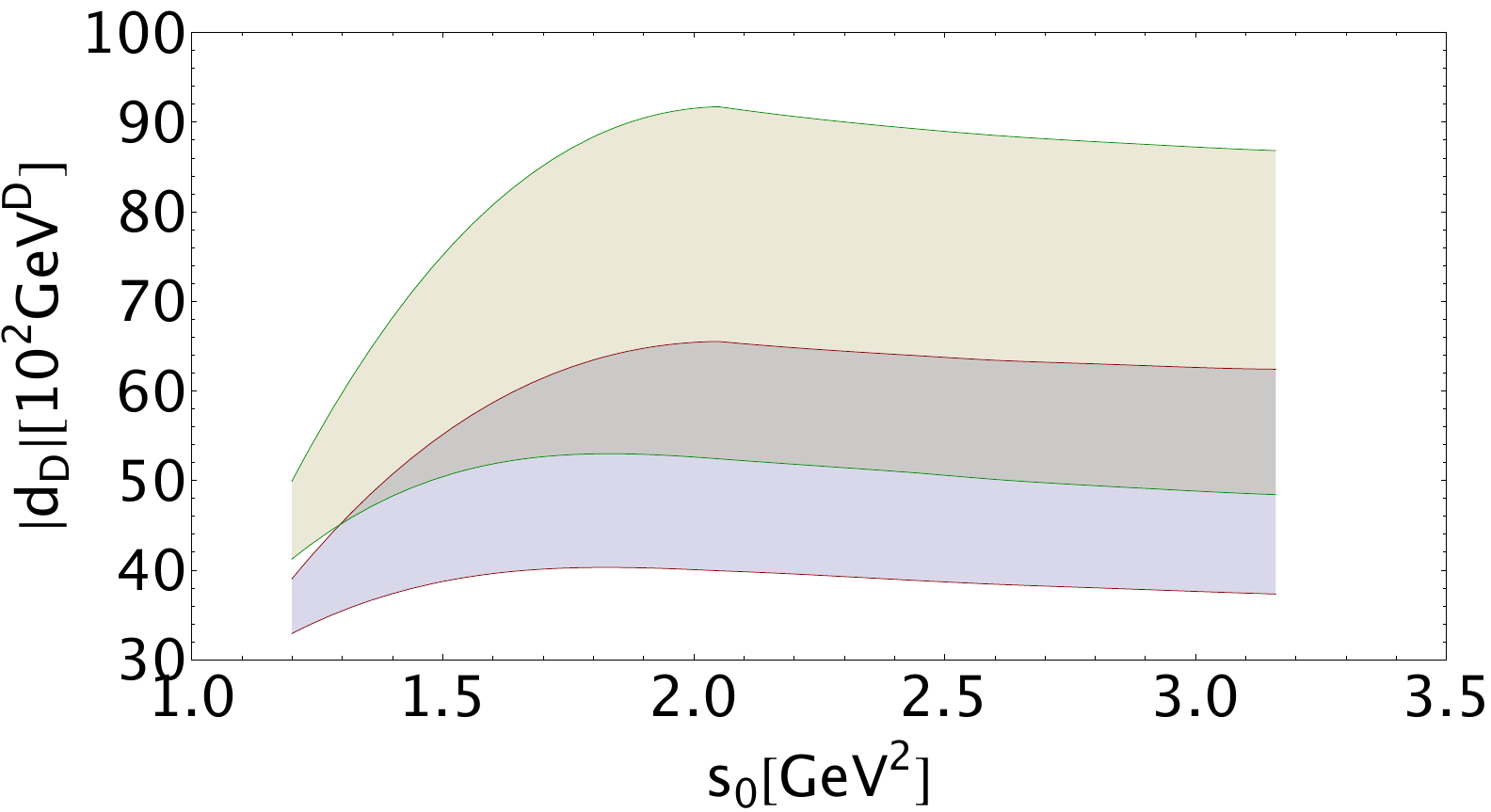}
\caption{\footnotesize  {\bf a)}: $|d_{6,A}|$ (lowest curve) and $|d_{8,A}|$ (highest curve)  versus  $M^2_0$; {\bf b)}: similar to  {\bf a)} but for $|d_{8,A}|$ and $|d_{10,A}|$. } \label{fig:d6-8}
\end{center}
\vspace*{-0.5cm}
\end{figure} 
We confront the experimental and QCD sides of ${\cal R}_{0,A}$ for different values of $s_0$. Like in the case of the ratio of Laplace sum rules,  we use a two-parameter fit ($d_{6,A},d_{8,A}$) to extract the values of these condensates using as input the values of the $d_4$ condensates and light quark masses. 

The results versus $s_0$ are shown in Fig.\,\ref{fig:d6-8}a). 
We consider as a reliable value the one from $s_0 = 1.93$ GeV$^2$ beyond the peak of the $A_1$ meson where a stability (minimum) around 3 GeV$^2$ just below the physical $\tau$ mass is found, We consider as our optimal value at this scale:
\bea
d_{6,A} &=&  (36.2\pm 0.5)\times 10^{-2}\,{\rm GeV^6}; ~ ~ d_{8,A} = -(55.2\pm 0.9)\times 10^{-2}\,{\rm GeV^8} ~    {\rm (FO)} \nnb\\
d_{6,A}  &=&  (33.1\pm 0.5)\times 10^{-2}\,{\rm GeV^6} ;~ ~ d_{8,A} = -(50.6\pm 0.8)\times 10^{-2}\,{\rm GeV^8}. ~    {\rm (CI)} \nnb\\
\eea

 One can notice from the analysis that the absolute values of the condensates are slightly higher for fixed order (FO)\,\cite{BNP,BNP2} than for contour improved (CI)\,\cite{LEDI} PT series. To be conservative, we take the arithmetic average of the FO  and CI values and add as a systematic the largest distance between the mean and the individual value\,:
\beq
d_{6,A} =  (34.6\pm1.8)\times 10^{-2}\,{\rm GeV^6} ~ ~~~~~~~~~~~~~~ d_{8,A} = -(52.9\pm2.4)\times 10^{-2}\,{\rm GeV^8}.
\label{eq:d68}
\eeq   

\section{ The dimension-eight and -ten condensates from ${\cal R}_{1,A}$}  
The expression of ${\cal R}_{1,A}$ is similar to ${\cal R}_{1,V}$ given in Eq. 20 of Ref.\,\cite{SNe}.  Using a two-parameter fit  ($d_{8,A}, d_{10,A}$) of ${\cal R}_{1,A}$ for different $s_0$, we show the result
of the analysis in Fig.\,\ref{fig:d6-8}b) using FO  PT series.   We obtain:
\beq
d_{8,A} =  -(51.4\pm 11.0)\times 10^{-2}\,{\rm GeV^8} ~ ~~~~~~~ d_{10,A} = (70.1\pm 21.6)\times 10^{-2}\,{\rm GeV^{10}}.
\label{eq:d810}
\eeq
 The results are stable versus $s_0$ but less accurate than in the case of ${\cal R}_{0,A}$ such that one cannot differentiate a FO from CI truncation of the PT series. Then, for higher moments, we shall consider , for definiteness, FO PT series.

\section{Final values of $d_{6,A}$ and $d_{8,A}$ from ${\cal R}_{0,A}$ and ${\cal R}_{1,A}$}
As a  final value of $d_{6,A}$, we take the mean of the ones in Eqs.\,\ref{eq:d68-ratio} and \ref{eq:d68} while for  $d_{8,A}$, we take the mean of the ones obtained in Eqs. \,\ref{eq:d68-ratio}, \ref{eq:d68} and \ref{eq:d810}. These values are quoted in Table\,\ref{tab:other}. 

\d We notice that the relation\,:
$
d_{6,A}\simeq  -(11/7)\,d_{6,V},
$
 is quite well satisfied.
 
 \d This result also suggests a violation of the four-quark condensate vacuum saturation 
 similar to the one found from $e^+e^-\to$ Hadrons data\,\cite{SNe,SNe2}\,:
 \beq
 \rho\alpha_s\la\bar \psi\psi\ra^2 = (6.38\pm 0.30)\times 10^{-4}\,{\rm GeV}^6~~~~ \lrar2 ~~~~ \rho\simeq  (6.38\pm 0.30),
 \eeq
 where $\rho=1$ if one would have used the vacuum saturation estimate  of the four-quark condensate:
 \beq
 \alpha_s\la\bar \psi\psi\ra^2 \simeq 1\times 10^{-4}\,{\rm GeV}^6.
 \eeq
\section{$\alpha_s(M_\tau)$ from the lowest BNP moment ${\cal R}_{0,A}$}
\begin{figure}[hbt]
\begin{center}
\includegraphics[width=10cm]{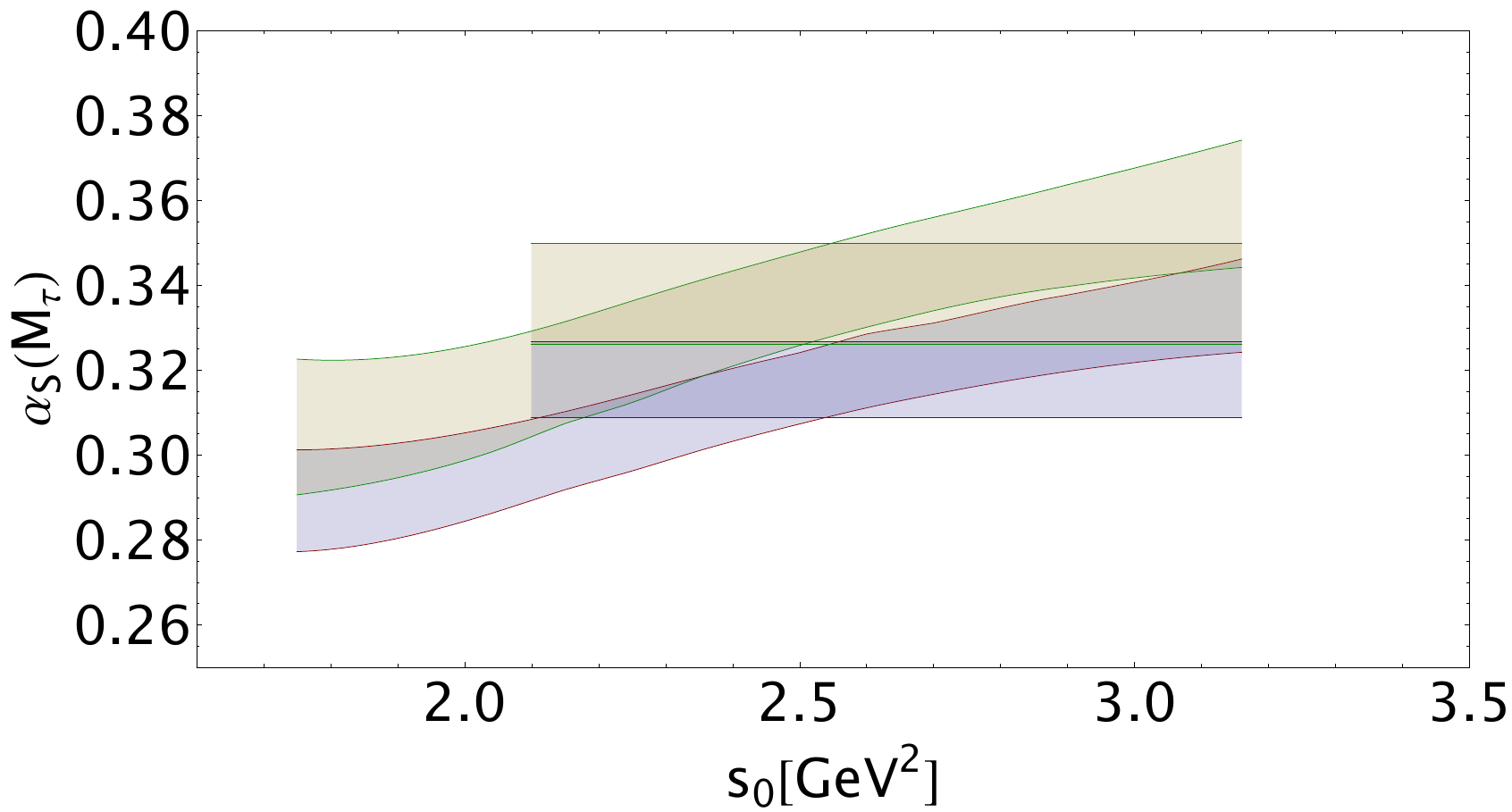}
\caption{\footnotesize  $\alpha_s(M_\tau)$  versus the hypothetical $\tau$-mass squared $s_0$. } \label{fig:as-A}
\end{center}
\vspace*{-0.5cm}
\end{figure}

We use the previous values of $d_{6,A} $ and $d_{8,A}$ together with the one of $\la\alpha_s G^2\ra$ in Eq.\ref{eq:g2} as inputs in ${\cal R}_{0,A}$ for extracting $\alpha_s(M_\tau)$. We show in Fig.\,\ref{fig:as-A}, the behaviour of $\alpha_s(M_\tau)$ versus an hypothetical $\tau$ mass squared $M_0^2\equiv s_0$. One can notice an inflexion point in the region $2.5^{+0.10}_{-0.15}$ GeV$^2$ at which we extract the optimal result. However, we shall consider the conservative result
from $s_0=2.1$ GeV$^2$ to $M_\tau^2$ (see Fig.\,\ref{fig:as-A}) which is\,:
\bea
\alpha_s(M_\tau)\vert_A &=&  0.3178(10) (65)~\lrar2~\alpha_s(M_Z)\vert_A =  0.1182(8)(3)_{evol} ~    {\rm (FO)} \nnb\\
&=& 0.3380(10)(43)~ \lrar2~ \alpha_s(M_Z)\vert_A =  0.1206(5)(3)_{evol} ~   {\rm (CI)}.
\label{eq:as-A}
 \eea
 The error in $\alpha_s(M_\tau)\vert_A$ comes respectively from the fitting procedure and from  an estimate of the $\alpha_s^5$ contribution from Ref.\,\cite{SNe}.  At the scale $s_0=$2.5 GeV$^2$ the sum of non-perturbative contributions to the moment normalized to the parton model is:
 \beq
 \delta_{NP,A}\simeq -(7.9\pm 1.1) \times 10^{-2}.
 \eeq
One can notice that extracting $\alpha_s(M_\tau)\vert_A $ at the observed $\tau$-mass tends to overestimate the result:
 \beq
 \alpha_s(M_\tau)\vert_A = 0.3352 (40)~~~{\rm FO},  ~~~~~~~~~~ 0.3592(47)~~~{\rm CI}.
 \label{eq:as-tau-A}
 \eeq
 These values extracted at $M_\tau$ agree with the ones from Ref.\,\cite{PICH1} obtained at the same scale. The same feature has been observed in the case of $e^+e^-\to$ Hadrons. 

\section{Comparison of $d_{6,A}$, $d_{8,A}$ and $\alpha_s(M_\tau)$ with previous results}
 \vspace*{-0.5cm}
   {\scriptsize
   \begin{center}
\begin{table}[hbt]
\setlength{\tabcolsep}{0.9pc}
  \begin{center}
    {\footnotesize
  \begin{tabular}{ccc cc ll}

&\\
\hline
\hline
\oliva Channel  &\oliva$d_6$ &\oliva$-d_8$ &\oliva$\alpha_s(M_\tau)$ FO&\oliva$\alpha_s(M_\tau)$ CI &\oliva Refs.\\
 \hline 
$e^+e^-$&$-(26.3\pm 3.7)$&$-(18.2\pm 0.6)$&0.3081(86)&0.3260(78)& \cite{SNe}\\
V&&&0.3129(79)&0.3291(70) &\cite{SNe}\\ 
\hline
A&$34.4\pm 1.7$&$51.5\pm 2.1$&0.3157(65)&0.3368(45)&This work\,\\
 A&$43.4\pm 13.8$&$59.2\pm 19.7$&$0.3390(180)$&$0.3640(230)$&\,\cite{PICH1}\\
A & $19.7\pm 1.0$&$27\pm 1.2$&{--}&0.3350(120)& \,\cite{ALEPH} \\
A& $9.6\pm 3.3$&$9.0\pm 5.0$&0.3230(160)&0.3470(230) &\,\cite{OPAL} \\

   \hline\hline
\end{tabular}}
 \caption{Values of the QCD condensates from some other $\tau$-moments at Fixed Order (FO) PT series and of $\alpha_s(M_\tau)$ for FO and Contour Improved (CI) PT series.}\label{tab:other} 
 \end{center}
\end{table}
\end{center}
} 
We compare our results with the ones from $e^+e^-$ and $\tau$-decay Vector channel\,\cite{SNe} and with the results obtained by different
authors in the Axial-Vector channel:

\d Our values of  $d_{6,A}$ and $d_{8,A}$ are in good agreement within the errors with the ones of Ref.\,\cite{PICH1} but about two times larger than the ones of Ref.\cite{ALEPH}. 

\d The value of $d_{8,A}$ suggests that the assumption\,: 
$d_{8,A}\approx d_{8,V}$
 is not satisfied by the fitted values given in Table\,\ref{tab:other}. 
\section{ High-dimension condensates from ${\cal R}_{n,A}$ }  
To determine the high-dimension condensates,  we  use the analogue of the moments given in Eq. 19 of Ref.\,\cite{SNe} for the vector channel and use as in the previous case the $s_0$-stability criteria for extracting the optimal result.
\subsection*{\b $d_{10,A}$ and $d_{12,A}$ condensates from ${\cal R}_{2,A}$}  
We  use a two-parameter fit to extract $(d_{10,A},d_{12,A})$ from    ${\cal R}_{2,A}$.
The $s_0$ behaviour of the results  in given in \,\cite{SNtau24}.
We deduce the values given in Table\,\ref{tab:cond-A}. 
\subsection*{\b $d_{2n,A}$ and $d_{2(n+1),A}$ condensates from higher moments}  
We do the same procedure as previously for higher dimension condensates $n\geq 6$.  The $s_0$  behaviours of the 
condensates are shown explicitly in\,\cite{SNtau24} where $s_0$-stability is found from 1.9 to $M_\tau^2$. The results  are summarized in Table\,\ref{tab:cond-A}.
 
\d One can notice that  the condensates in the A channel has alternate signs, while 
 their absolute size is almost constant and more accurate than previous determinations in the literature. 
 
 \d There is not also any indication of an exponential growth. According to Ref.\,\cite{SHIFMAN}, this feature in the Euclidian region does not favour  a sizeable duality violation  (DV) in the time-like one\,\cite{DV}. 

   {
\begin{table}[H]
\setlength{\tabcolsep}{0.1pc}
  \begin{center}
    {\footnotesize
  \begin{tabular}{lllll ll ll}

&\\
\hline
\hline
$\oliva d_{6,A}$&$\oliva -d_{8,A}$&$\oliva d_{10,A}$&$\oliva -d_{12,A}$&$\oliva d_{14,A}$& $\oliva -d_{16,A}$&$\oliva d_{18,A}$&$\oliva- d_{20,A}$&Refs.\\
 \hline 
$34.4\pm 1.7$&$51.5\pm 2.1$&$53.9\pm 0.7$&$63.3\pm1.8$&$77.0\pm 6.0$&$93.1\pm 4.0$&$104.3\pm 7.7$&$ 119.7\pm 8.2$& This work\\
$43.4\pm 13.8$&$59.2\pm 19.7$&$63.2\pm 33.6$&$43.4\pm31.6$&&&&&\cite{PICH1}\\
 $19.7\pm 1.0$&$27\pm 1.2$&&&&&&&\cite{ALEPH} \\
 $9.6\pm 3.3$&$9.0\pm 5.0$&&&&&&&\cite{OPAL} \\
   \hline\hline
\end{tabular}}
 \caption{ Values of the QCD condensates of dimension $D$ in units of $10^{-2}$ GeV$^{D}$ from  this work and some other estimates. }\label{tab:cond-A} 
 \end{center}
\end{table}
} 
\section{$\alpha_s(M_\tau)$ from the lowest BNP moment ${\cal R}_{0,V-A}$ } 
In the chiral limit their expression is similar to the previous ones for the axial-vector current modulo an overall factor 2 and the change into the condensates contributions $d_{D,V-A}$ of dimension $D$.

The $s_0$ behaviour of the moment is shown explicitly in Ref.\,\cite{SNtau24}.  
For  $s_0=M_\tau^2$, we obtain :
${\cal R}_{0,V-A}= 3.484 \pm 0.022$
to be compared with the ALEPH data\,\cite{ALEPH}:
$
{\cal R}_{0,V-A}\vert_{Aleph} = 3.475\pm 0.011,
$
where our error is larger as  we have  separately fitted the upper and lower values of the data. 
\begin{figure}[hbt]
\begin{center}
\includegraphics[width=10cm]{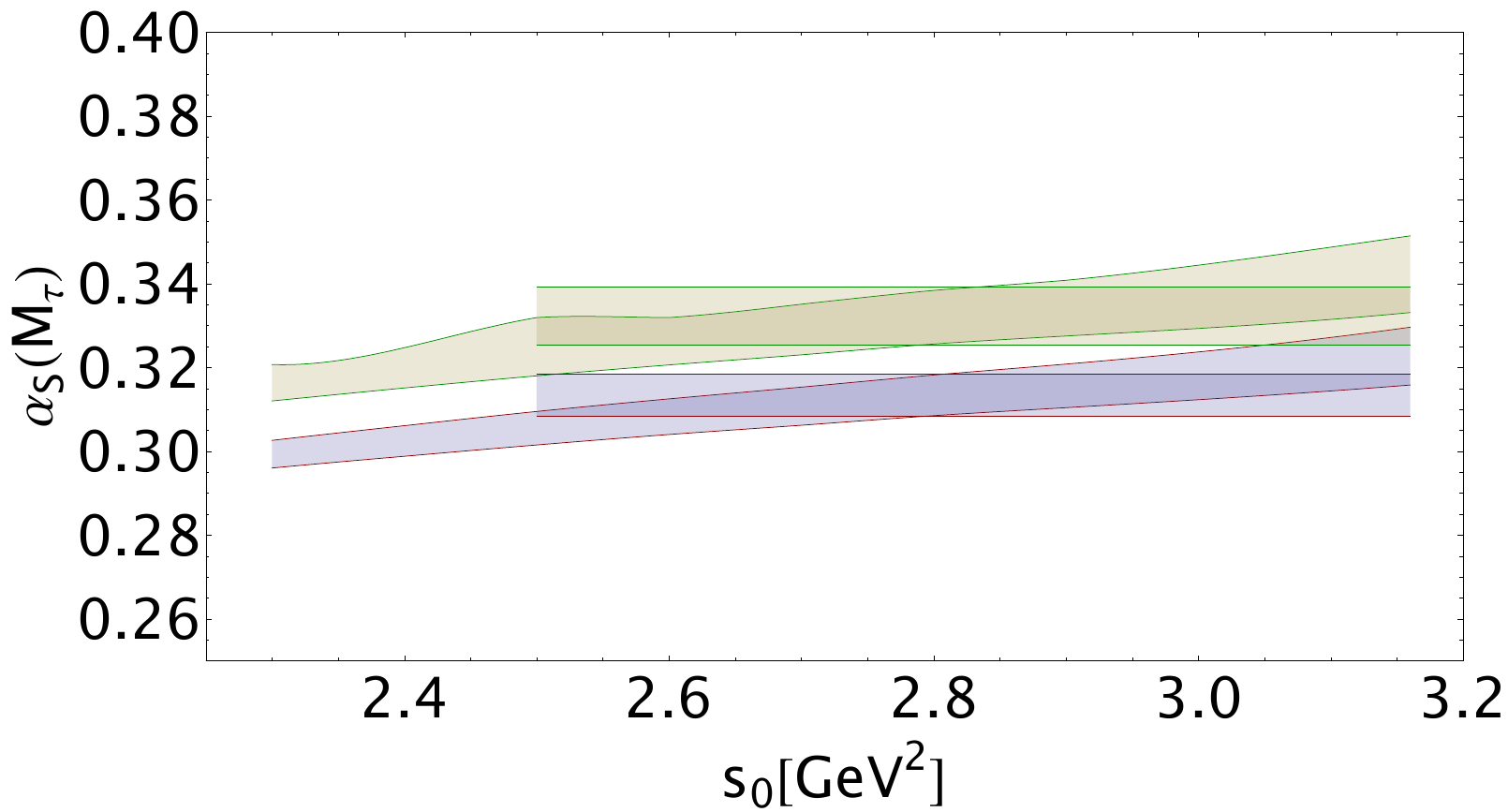}
\caption{\footnotesize  $\alpha_s(M_\tau)$  versus an hypothetical $\tau$-mass squared $s_0$. The upper curves corresponds to CI perturbative series and the lower ones to FO. The horizontal lines come from a least-square fit of the data in the optimal region $s_0\simeq (2.3\sim2.9)$ GeV$^2$.} \label{fig:as-VA}
\end{center}
\vspace*{-0.5cm}
\end{figure} 

Using as inputs the value of $\la\alpha_s G^2\ra$ in Eq.\,\ref{eq:g2} and the 
ones\,\cite{SNtau24}\,:
\beq
d_{D,V-A} \equiv \frac{1}{2}\ga d_{D,V}+d_{D,A}\dr\,:~~~~~~D=6,8,
\label{eq:d68-VA}
\eeq
we extract the value of $\alpha_s(M_\tau)$ as a function of $s_0$ as shown in  Fig.\,\ref{fig:as-VA}) from ${\cal R}_{0,V-A}(s_0)$. We notice a stable result in the region $s_0\simeq (2.5\sim 2.6)$ GeV$^2$ though not quite convincing. The, we consider as a conservative value the one obtained from a least-square fit of the values inside the region  $[2.5,M_\tau^2]$. The optimal result corresponds to $s_0=2.8$ GeV$^2$ (see Fig.\,\ref{fig:as-VA}):
\bea
\alpha_s(M_\tau)\vert_{V-A} &=&  0.3135(51) (65)~\lrar2~ \alpha_s(M_Z)\vert_{V-A} =  0.1177(10)(3)_{evol} ~   {\rm (FO)} \nnb\\
&=& 0.3322(69)(43)~ \lrar2~\alpha_s(M_Z)\vert_{V-A} =  0.1200(9)(3)_{evol} ~    {\rm (CI)}.\nnb\\
\label{eq:as-VA}
 \eea
 
 The  errors in $\alpha_s(M_\tau)\vert_{V-A}$ comes respectively from the fitting procedure and from  an estimate of the $\alpha_s^5$ contribution from Ref.\,\cite{SNe}.  At this scale the sum of non-perturbative contributions to the moment normalized to the parton model is:
 \beq
 \delta_{NP,V-A}\simeq +(2.7\pm 1.1) \times 10^{-4},
 \eeq
 which is completely negligible.
One can notice from Fig.\,\ref{fig:as-VA} that extracting $\alpha_s(M_\tau)\vert_{V-A} $ at the observed $M_\tau$-mass also tends to overestimate its value\,:
 \beq
 \alpha_s(M_\tau)\vert_{V-A}= 0.3227 (69)(65)~~~{\rm FO},  ~~~~~~~~~~ 0.3423(92)(43)~~~{\rm CI}
 \label{eq:as-tau-VA}
 \eeq
 like in the case of the axial-vector channel and $e^+e^-\to$ Hadrons data.

\subsection*{\b Comparison with some previous results}
   {\scriptsize
   \begin{center}
\begin{table}[hbt]
\setlength{\tabcolsep}{0.9pc}
  \begin{center}
    {\footnotesize
  \begin{tabular}{ccc cc ll}

&\\
\hline
\hline
\oliva$d_{6,V-A}$ &\oliva$-d_{8,V-A}$&\oliva$\alpha_s(M_\tau)$ FO&\oliva$\alpha_s(M_\tau)$ CI  & \oliva $s_0$ [GeV]$^2$&\oliva Refs.\\
 \hline 
$3.6\pm 0.5$&$14.5\pm 2.2$&0.3135(83)&0.3322(81)& 2.5 $\to M_\tau^2$&This work\,\\
\hline
$3.6\pm 0.5$&$14.5\pm 2.2$&0.3227(95)&0.3423(102)& $M_\tau^2$&This work\,\\
 $5.1^{+5.5}_{-3.1}$&$3.2\pm 2.2$&$0.3170^{+0.0100}_{-0.0050}$&$ 0.3360^{+0.0110}_{- 0.0090}$&$M_\tau^2\,(D\leq10)$ &Table 7 \,\cite{PICH1}\\
  $24^{+24}_{-16}$&$32^{+25}_{-32}$&$0.3290^{+0.0120}_{-0.0110}$&$ 0.3490^{+0.0160}_{- 0.0140}$&$M_\tau^2\,(D\leq12)$ &Table 8 \,\cite{PICH1} \\

  $2.4\pm 0.8$&$3.2\pm 0.8$&--&$0.3410(78)$&$M_\tau^2$&\,\cite{ALEPH}\\
    $1.5\pm 4.8$&$3.7\pm 9.0$&0.3240(145)&$0.3480(212)$&$M_\tau^2$&\,\cite{OPAL}\\
    
   \hline\hline
\end{tabular}}
 \caption{Values of the QCD condensates from ${\cal R}_{0,e^+e^-}$ and ${\cal R}_{0,A}$ at Fixed Order (FO) PT series and of $\alpha_s(M_\tau)$ for FO and Contour Improved (CI) PT series. The condensates are in units of $10^{-2}$ GeV$^D$.}\label{tab:other-VA} 
 \end{center}
\end{table}
\end{center}
} 
In Table\,\ref{tab:other-VA}, we compare our results with some other determinations\,\cite{ALEPH,PICH1,OPAL}\,\footnote{Some estimates including renormalon within a large $\beta$ approximation [resp. duality violation] effects can be e.g. found  in Refs.\,\cite{CVETIC}  [resp. \cite{DV}].}  obtained at the scale $s_0=M_\tau^2$.  

\d There is a quite good agreement for $d_{6,V-A}$ but not for $d_{8,V-A}$ from different determinations.


\d Extracting  $\alpha_s$ at $M_\tau$, there is a good agreement among different determinations where the values are slightly higher than the ones from the optimal region  given in the first row\, (see Fig.\,\ref{fig:as-VA}). 
\section{Mean value of $\alpha_s$ from $e^+e^-\to$ Hadrons and  $\tau$-decays}  
Using the result from $e^+e^-\to$ Hadrons and from the A and V--A $\tau$-decay channels, we deduce the mean:
\bea
 \alpha_s(M_\tau)&=&0.3140(44)\, {\rm (FO)} ~ ~~~\lrar2~ ~~~ \alpha_s(M_Z) = 1178(6)_{fit}(3)_{evol.},\nnb\\
 &=& 0.3346 (35) \, {\rm (CI)} ~ ~~~\lrar2~ ~~~\alpha_s(M_Z)=0.1202(4)_{fit}(3)_{evol.}.
 \label{eq:as-mean}
 \eea
 This result for FO is in perfect agreement with the PDG world average without lattice results\,\cite{PDG} while the one for CI is slightly larger than the one including lattice ones. 
\section{Conclusion}  
\d We have summarized the determinations of the QCD condensates in the axial-vector (A) and V--A channels done in Ref.\,\cite{SNtau24}  from the  ratio of LSR and higher BNP-like moments  (see Tables\,\ref{tab:cond-A} and \ref{tab:other-VA}). We observe alternate signs in the A channel and an almost constant value of their absolute size for the A and V--A channels. The absence of an exponential behaviour in the Euclidian region may not favour a duality violation in the time-like region\,\cite{SHIFMAN}. 

\d Using as inputs the value of the QCD condensates up to $d_{8}$, we extract $\alpha_s(M_\tau)$ from the lowest BNP-moment ${\cal R}_{0,A} (s_0)$ and ${\cal R}_{0,V-A} (s_0)$. The results from the A and V-A channels are 
 given in Eqs.\,\ref{eq:as-A} and \ref{eq:as-VA} while their mean with the one from the $e^+e^-\to$ Hadrons data is given in Eq.\,\ref{eq:as-mean}. Our results are in perfect agreement with the PDG world averge\,\cite{PDG}.

\d We have not considered in our analysis some eventual effects beyond the SVZ-expansion which can be neglected as discussed in Section\,\ref{sec:beyond}.
\section*{Declaration of competing interest}  
The author declares that he has no known competing financial interests or personal relationships that could have appeared to influence the work reported in this paper.
\section*{Data availability}  
No data was used for the research described in the article.


\begin{thebibliography}{999}
\vspace*{-0.15cm}
 \bibitem{SNtau24} S. Narison,  {\it Nucl. Phys.}{ \bf A1055} (2025) 123014. 
\bibitem{SNe} S. Narison, 
{\it Nucl. Phys.}{ \bf A 1046} (2024) 122873, {\it Nucl. Phys.} {\bf A1050} (2024) 122915 (erratum);
{\it Nucl. Part. Phys. Proc.}. {\bf 347} (2024) 105.
\bibitem{SNe2} S. Narison, {\it Nucl. Phys.} {\bf A 1039} (2023) 122744; 
{\it Nucl. Part. Phys. Proc.} {\bf 343} (2024). 
\bibitem{SVZ}M.A. Shifman, A.I. Vainshtein, V.I. Zakharov, {\it Nucl. Phys.} {\bf B147} (1979) 385; 
{\it Nucl. Phys.} {\bf B147} (1979) 448.

\bibitem{BNP}E. Braaten, S. Narison, A. Pich, {\it Nucl. Phys.} {\bf B373} (1992) 581.

\bibitem{BNP2}E. Braaten, {\it Phys. Rev. Lett.} {\bf 60} (1988) 606; E. Braaten, {\it Phys. Rev.} {\bf D39} (1989) 1458; 
S. Narison, A. Pich, {\it Phys. Lett.} {\bf B211} (1988) 183.

%



\bibitem{SNB2}For a review, see e.g. S. Narison, {\it QCD as a theory of hadrons, Cambridge Monogr. Part. Phys. Nucl. Phys. Cosmol.} {\bf 17}.
 (2004) 1-778 [hep-ph/0205006]. 
 




\bibitem{PDG}R.L. Workman et al. (Particle Data Group), {\it Prog. Theor. Exp. Phys.}  {\bf 2022} (2022) 083C01.

\bibitem{ALEPH} The ALEPH collaboration: M. Davier, A. Hoecker, B. Malaescu, C. Yuan, Z. Zhang, {\it Eur. Phys. J.} {\bf  C74}  (2014) 2803.

\bibitem{SNR}S. Narison and E. de Rafael,  {\it Phys. Lett.} {\bf B 103},  (1981) 57.
\bibitem{BELL}J.S. Bell and R.A. Bertlmann, {\it Nucl. Phys.} {\bf B187}, (1981) 285.


\bibitem{SNLSR} 
S. Narison,  {\it The Laplace Transform and its Applications}, ed. V. Martinez-Lucaes, Nova Science Pub., New-York - 2024 (arXiv: 2309.00258 [hep-ph]).

  \bibitem{SNparam}S. Narison,  {\it Int. J. Mod. Phys.} {\bf A33} (2018) no. 10, 185004; 
  Addendum: {\it Int. J. Mod. Phys.} {\bf A33} (2018) no.10, 1850045.
\bibitem{SNcb1}S. Narison,  {\it Phys. Lett.} {\bf B693}  (2010) 559, erratum {\it ibid}, 
{\bf B705} (2011) 544; 
ibid, {\bf B706}  (2012) 412; 
ibid, {\bf B707}  (2012) 259.


 \bibitem{LEDI}F. Le Diberder,  A. Pich, {\it Phys. Lett.} {\bf B286} (1992) 147; {\bf B289} (1992) 165.

\bibitem{SNZ} S. Narison, V.I. Zakharov, {\it Phys. Lett.} {\bf B679} (2009) 355.

\bibitem{CNZ} K.G. Chetyrkin, S. Narison, V. I. Zakharov, {\it Nucl. Phys.} {\bf B550}(1999) 353. 

\bibitem{CVETIC} C. Ayala, G. Cvetic, D. Teca, {\it J.Phys.} {\bf G 50} (2023) $n^0.$ 4, 045004; 
M. Beneke, D. Boito,  M. Jamin,   {\it JHEP} {\bf 01} (2013) 125.

\bibitem{DV} D. Boito, M. Golterman, K. Maltman, S. Peris, {\it Phys.
Rev.} {\bf D 95} (2017) $n^o$. 3, 034024.

\bibitem{PICH1} A. Pich, A. Rodr\'\i guez-S\'anchez, {\it Phys.Rev.} {\bf D94} (2016) 3, 034027;  {\it Mod.Phys.Lett.} {\bf A31} (2016) 30, 1630032; {\it Nucl. Part. Phys. Proc.}{\bf 287-288} (2017) 81.

\bibitem{SHIFMAN} M.A. Shifman, {\it Nucl. Phys. Proc. Suppl.}  {\bf B207-208} (2010) 298; ibid arXiv: hep-ph/0009131; 
B. Blok, M.A. Shifman,  Da-Xin Zhang, {\it Phys. Rev.} {\bf D 57} (1998) 2691; {\it Phys. Rev.} {\bf D 59} (1999) 019901 (erratum). 

\bibitem{OPAL} The OPAL collaboration: K. Ackerstaff et al., {\it Euro. Phys. J.} {\bf C7} (1999) 571. 



\end{thebibliography}
\end{document}